\documentclass[11pt, twoside, eqno]{article}
\usepackage{amssymb}
\usepackage{mathrsfs}
\usepackage{amsmath}
\usepackage{amsfonts}
\usepackage{amsthm}
\usepackage{subfigure}
\usepackage{graphicx}
\usepackage{booktabs}
\usepackage{setspace}
\usepackage{float}

\allowdisplaybreaks \numberwithin{equation}{section}

\def\hs{\hspace{0.5cm}}

\begin{document}
\baselineskip=15pt
\renewcommand{\arraystretch}{2}
\arraycolsep=1pt
\title{\bf \Large Zero-Correlation Linear Cryptanalysis of Reduced Round  ARIA  with  Partial-sum  and FFT
\footnotetext{\hspace{-0.6cm} ${ }^{*}$ Corresponding authors. \\
   E-mail addresses: nlwt8988@gmail.com.}
\author{\vspace{-0.1cm}\bf  Wentan Yi$^{*}$, Shaozhen Chen and Kuanyang Wei \\
\vspace{-0.5cm}\small\it State Key Laboratory of Mathematical Engineering and Advanced Computing,\\
\small\it Zhengzhou 450001, China }}
\date{}

\maketitle

\begin{center}
\begin{minipage}{15.2cm}
\small{\bf Abstract.} Block cipher ARIA was first proposed by some South Korean experts in 2003, and later, it was established as a Korean Standard block cipher algorithm by Korean Agency for Technology and Standards.  In this paper, we focus on the security evaluation of ARIA block cipher against the recent zero-correlation linear cryptanalysis. In addition, Partial-sum technique and  FFT (Fast Fourier Transform) technique are used to speed up the cryptanalysis, respectively.

\quad We first introduce some 4-round linear approximations of ARIA with zero-correlation, and then present some key-recovery attacks on 6/7-round ARIA-128/256 with Partial-sum technique and FFT technique.
The key-recovery attack with Partial-sum technique on
6-round ARIA-128 needs $2^{123.6}$ known plaintexts (KPs), $2^{121}$ encryptions and $2^{90.3}$ bytes memory,
and the attack with FFT technique requires $2^{124.1}$ KPs, $2^{121.5}$ encryptions and $2^{90.3}$ bytes memory. Moreover,  applying Partial-sum technique, we can attack 7-round ARIA-256 with $2^{124.6}$ KPs, $2^{203.5}$ encryptions and $2^{152}$ bytes memory and 7-round ARIA-256 employing FFT technique, requires $2^{124.7}$ KPs, $2^{209.5}$ encryptions and $2^{152}$ bytes memory. Our results are the first zero-correlation linear cryptanalysis results on ARIA.
\medskip

\end{minipage}
\end{center}

\section{\large\bf Introduction}

ARIA [1] is a  block cipher designed by a group of Korean experts in 2003. In 2004, ARIA
was established as a Korean Standard block cipher algorithm by the Ministry of Commerce, Industry and Energy.
ARIA is a general-purpose involutional SPN(substitution permutation network) block cipher algorithm, optimized for both lightweight environments and hardware implementation. ARIA supports 128-bit block length with the key sizes of 128/192/256 bits, and the
most interesting characteristic is its involution based on the special usage of
neighbouring confusion layer and involutional diffusion layer.

The security of ARIA has been internally evaluated by the designers [1] with differential cryptanalysis, linear cryptanalysis, truncated differential cryptanalysis, impossible differential cryptanalysis, higher order differential cryptanalysis,
square attack and interpolation attack. Biryukov et al.[2] performed an evaluation of ARIA with truncated differential cryptanalysis and dedicated linear cryptanalysis. For the first time, Wu et al. [3] found a non-trivial 4-round
impossible differentials and they gave a attack on 6-round ARIA requiring
about $2^{121}$ chosen plaintexts and $2^{112}$ encryptions. Based on some properties of the binary matrix used in the diffusion layer, Li et al.[4] found some new 4-round impossible differentials of ARIA, and they gave an efficient attack on 6-round ARIA.
Later, Fleischmann et al.[5] proposed the boomerang attack on 6-round ARIA and the integral attack [6] was introduced in the analysis of 7-round ARIA. Tang et al.[7]  proposed the  meet-in-the-middle attack on 7-round ARIA.
Du et al.[8] proposed the impossible differentials  on 7-round ARIA-256 and recently, Xie et al.[9]  gave some improvements. Attack results on ARIA are summarized in Table 1.

\begin{table}[tbp]
\centering
\scriptsize
\begin{tabular}{cccccc}
\hline
Attack Type &key size& Rounds & Date & Time & Source\\
\hline
Truncated Differential & 128 & 7 &$2^{81}$ CPs & $2^{81}$ Enc & [2] \\
\hline

\vspace{-0.12in}Impossible Differential & 128 & 6 &$2^{121}$ CPs & $2^{121}$ Enc & [3] \\
\vspace{-0.12in} Impossible Differential  & 192 & 7&$2^{127}$ CPs & $2^{176.2}$ Enc & [9] \\
Impossible Differential  & 256 & 7&$2^{125}$ CPs & $2^{238}$ Enc & [8]\\

\hline
Meet-in-Middle  & 192 & 7 &$2^{120}$ KPs & $2^{185.3}$ Enc & [7]\\
\hline
Boombrang & 192 & 6 &$2^{57}$ CPs & $2^{171.2}$ Enc & [5] \\
\hline
\vspace{-0.12in}Integral & 128 & 6 &$2^{99.2}$ CPs & $2^{74.1}$ Enc & [6] \\
Integral & 256 & 7 &$2^{100.6}$ CPs & $2^{225.8}$ Enc & [6] \\
\hline
\vspace{-0.12in} ZC.Partial-sum   & 128 & 6 &$2^{123.6}$ KPs & $2^{121}$ Enc & Sect.4.1\\
\vspace{-0.12in}ZC.FFT   & 128 & 6 &$2^{124.1}$ KPs & $2^{121.5}$ Enc & Sect.4.2\\
\vspace{-0.12in}ZC.Partial-sum   & 256 & 7 &$2^{124.6}$ KPs & $2^{203.5}$ Enc & Sect.5.1\\
ZC.FFT   & 256 & 7 &$2^{124.7}$ KPs & $2^{209.5}$ Enc & Sect.5.2\\
\hline
\end{tabular}

KP(CP) refer to the number of  known(chosen) plaintexts,
 Enc refers to the number of encryptions.
\caption{Summary of the main attacks on ARIA}
\end{table}
Zero-correlation linear cryptanalysis [10] was  proposed by Bogdanov and Rijmen, which has its theoretical foundation in the availability of numerous key-independent unbiased linear approximations with correlation zero for many ciphers.
However, the initial distinguisher of [10] had some limitations in terms of data complexity, which needs at least half of the codebook.
In FSE 2012, Bogdanov and Wang [11] proposed a more data-efficient distinguisher by making use of  multiple linear approximations with zero-correlation. Although the date complexity is reduced, the distinguisher relies on the assumption that all linear approximations with zero-correlation are independent. At AsiaCrypt 2012[12], a multidimensional distinguisher has been constructed for the zero-correlation property, which removed the unnecessary independency assumptions. Recently, zero-correlation linear cryptanalysis has been using in the attack of block ciphers CAST-256[12], Camellia[13], CLEFIA[13], HIGHT[14],  LBlock[15] and E2[16], successfully.

Some improving techniques for zero-correlation linear cryptanalysis have been proposed, such as the partial-sum technique and FFT technique.
Ferguson et al. [17] proposed the partial-sum technique in 2000 and  they applied the technique to the integral attacks on 6-round AES.
The basic idea of Partial-sum technique is to  partially compute the sum by guessing each key one after another instead of guessing the all keys one time. Since zero-correlation linear cryptanalysis use enormous  plaintexts-ciphertexts pairs, Partial-sum technique can also be used to reduce the computation complexity in the attack procedure.

FFT-based technique of computational complexity reduction was first proposed by Collard et al.[18] in the linear attack on the AES candidate Serpent in 2007. It also relies on eliminating the redundant computations from the partial encryption/decryption in attack process. At SAC 2013,  Bogdanov et al.[13] applied FFT technique to the zero-correlation linear cryptanalysis of Camellia.

In this paper, 4-round zero-correlation linear approximations of ARIA are discussed in detail. Furthermore, we investigate the security
of 6/7-round ARIA-128/256 with both Partial-sum and FFT techniques. Our contributions can be summarized as follows.

1. We  reveal some 4-round zero-correlation linear approximations of ARIA. If we treat the input/output masks as the input/output differentials, they are 4-round impossible differentials of ARIA owing that the diffusion layer of the round function is a diagonal matrix.

2.  Based on those new linear approximations with zero-correlation, key-recovery attacks on 6/7-round  ARIA-128/256 are proposed.  In addition, we use Partial-sum technique and FFT technique to speed up the attacks.  To  my knowledge, they are the first zero-correlation linear attacks on reduced-round  ARIA.

The paper is organized as follows. Section 2 gives a brief description of block cipher ARIA and outlines the ideas of  zero-correlation linear cryptanalysis. Some new zero-correlation linear approximations are shown in Section 3. Section 4 and Section 5 illustrate our attacks on 6/7-round ARIA-128/256 with Partial-sum and FFT technique, respectively. We conclude in Section 6.

\section{\large \bf  Preliminarise}
\subsection{\bf Description of ARIA}

ARIA is an SPN  style block cipher and the number of the rounds are 12/14/16
corresponding to key of 128/192/256 bits. The round function constitutes 3 basic operations: the substitution
layer, the diffusion layer and the round key addition, which can be described as follows:

$\mathbf{Round\  Key \ Addition(KA):}$ This is done by XORing the 128-bit round key, which is derived from the cipher key by means of the key schedule.

$\mathbf{Substitution\  Layer(SL):}$ Applying the $8\times 8$ S-boxes 16 times in parallel on each byte.
There are two types of substitution layers to be used so as to make the cipher involution, see Figure 1.
For convenience, we denote by $S^{}_{r,k}$, $S^{-1}_{r,k}$ the $k$-th S-box of $r$-th round and its inverse S-box.
\begin {figure}[H]
\centering
 \includegraphics[width=10cm]{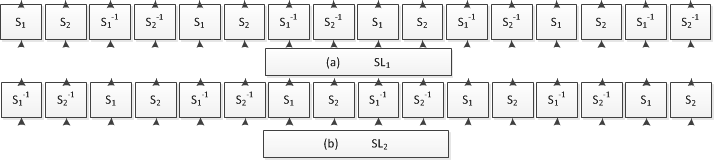}
  \caption{The substitution layer of ARIA}
\end {figure}


$\mathbf{Diffusion\  Layer(DL):}$ A linear map $P : (F^8_2 )^{16} \rightarrow (F^8_2 )^{16}$ is given by
$$(x_0,x_1,...,x_{15}) \rightarrow  (y_0,y_1,...,y_{15}),$$
where
\begin{equation*}
 \scriptsize
\left(\vspace{-0.2in}
\begin{array}{l}
 \vspace{-0.18in}y_0 \\
 \vspace{-0.18in}y_1 \\
 \vspace{-0.18in}y_2\\
 \vspace{-0.18in}y_3 \\
 \vspace{-0.18in}y_4 \\
 \vspace{-0.18in}y_5\\
 \vspace{-0.18in}y_6 \\
 \vspace{-0.18in}y_7\\
 \vspace{-0.18in}y_8 \\
 \vspace{-0.18in}y_9 \\
 \vspace{-0.18in}y_{10}\\
 \vspace{-0.18in}y_{11} \\
 \vspace{-0.18in}y_{12} \\
 \vspace{-0.18in}y_{13}\\
 \vspace{-0.18in}y_{14} \\
 y_{15}\\
\end{array}
\right) =\left(
\begin{array}{cccccccccccccccc}
 \vspace{-0.18in}0 & 0 & 0& 1 & 1 & 0 & 1 & 0 & 1 & 1 & 0& 0 & 0 & 1 & 1 & 0 \\
 \vspace{-0.18in}0 & 0 & 1& 0 & 0 & 1 & 0 & 1 & 1 & 1 & 0& 0 & 1 & 0 & 0 & 1 \\
 \vspace{-0.18in}0 & 1 & 0& 0 & 1 & 0 & 1 & 0 & 0 & 0 & 1& 1 & 1 & 0 & 0 & 1 \\
 \vspace{-0.18in}1 & 0 & 0& 0 & 0 & 1 & 0 & 1 & 0 & 0 & 1& 1 & 0 & 1 & 1 & 0 \\
 \vspace{-0.18in}1 & 0 & 1& 0 & 0 & 1 & 0 & 0 & 1 & 0 & 0& 1 & 0 & 0 & 1 & 1 \\
 \vspace{-0.18in}0 & 1 & 0& 1 & 1 & 0 & 0 & 0 & 0 & 1 & 1& 0 & 0 & 0 & 1 & 1 \\
 \vspace{-0.18in}1 & 0 & 1& 0 & 0 & 0 & 0 & 1 & 0 & 1 & 1& 0 & 1 & 1 & 0 & 0 \\
 \vspace{-0.18in}0 & 1 & 0& 1 & 0 & 0& 1 & 0 & 0 & 1 & 0& 1 & 1 & 1 & 0 & 0 \\
 \vspace{-0.18in}1 & 1 & 0& 0 & 1 & 0 & 0 & 1 & 0 & 0 & 1& 0 & 0 & 1 & 0 & 1 \\
 \vspace{-0.18in}1 & 1 & 0& 0 & 0 & 1 & 1 & 0 & 0 & 0 & 0& 1 & 1 & 0 & 1 & 0 \\
 \vspace{-0.18in}0 & 0 & 1& 1 & 0 & 1 & 1 & 0 & 1 & 0 & 0& 0 & 0 & 1 & 0 & 1 \\
 \vspace{-0.18in}0 & 0 & 1& 1 & 1 & 0 & 0 & 1 & 0 & 1 & 0& 0 & 1 & 0 & 1 & 0 \\
 \vspace{-0.18in}0 & 1 & 1& 0 & 0 & 0 & 1 & 1 & 0 & 1 & 0& 1 & 1 & 0 & 0 & 0 \\
 \vspace{-0.18in}1 & 0 & 0& 1 & 0 & 0 & 1 & 1 & 1 & 0 & 1& 0 & 0 & 1 & 0 & 0 \\
 \vspace{-0.18in}1 & 0 & 0& 1 & 1 & 1 & 0 & 0 & 0 & 1 & 0& 1 & 0 & 0 & 1 & 0 \\
 0 & 1 & 1& 0 & 1 & 1 & 0 & 0 & 1 & 0 & 1& 0 & 0 & 0 & 0 & 1 \\
\end{array}
\right)
\cdot
\left(
\begin{array}{ccc}
 \vspace{-0.18in}x_0 \\
 \vspace{-0.18in}x_1 \\
 \vspace{-0.18in}x_2\\
 \vspace{-0.18in}x_3 \\
 \vspace{-0.18in}x_4 \\
 \vspace{-0.18in}x_5\\
 \vspace{-0.18in}x_6 \\
 \vspace{-0.18in}x_7\\
 \vspace{-0.18in}x_8 \\
 \vspace{-0.18in}x_9 \\
 \vspace{-0.18in}x_{10}\\
 \vspace{-0.18in}x_{11} \\
 \vspace{-0.18in}x_{12} \\
 \vspace{-0.18in}x_{13}\\
 \vspace{-0.18in}x_{14} \\
 x_{15}\\
\end{array}
\right)
\end{equation*}

Note that the diffusion layer of the last round is replaced by a round key
addition. We shall assume that the 6/7-round ARIA also has the diffusion layer
replaced by a round key addition in the attack of 6/7-round ARIA. In addition, our attacks do not utilize the key relation, we omit the details of ARIA's key schedule.

\subsection{\bf Basic ideas of  zero-correlation linear cryptanalysis}

In this section, we briefly recall the basic concepts of zero-correlation linear cryptanalysis based on [10], [11] and [12]. Linear cryptanalysis is based on linear approximations determined by input mask $a$ and output mask $\beta$.  A linear approximation $a\rightarrow \beta$ of a vectorial function $f$ has a correlation denoted by
$$C(\beta \cdot f(x), a\cdot x)=2\textrm{Pr}_{x}(\beta \cdot f(x)\oplus a\cdot x=0)-1,$$
where we denote the scalar product of binary vectors by $a\cdot x = \oplus_{i=1}^{n}a_i x_i$.

In zero-correlation linear cryptanalysis, the distinguisher uses linear approximations with zero correlation for
all keys while the classical linear cryptanalysis utilizes linear approximations with correlation as far from zero as
possible. Zero-correlation linear cryptanalysis with multiple linear
approximations was introduced in [11].

Let the number of available zero-correlation linear approximations for an n-bit
block cipher be denoted by $l$. Let the number of required known plaintexts
be $N$. For each of the $l$ given linear approximations, the adversary computes
the number $T_i$ of times that linear approximation $i$ is fulfilled on $N$ plaintexts
and ciphertexts, $i \in \{ 1,...,l\}$. Each $T_i$ suggests an empirical correlation value
$\hat{c}_i = 2T_i/N-1$. Under a statistical independency assumption, $\sum_{i=0}^{l}\hat{c}^2_i$ follows a $\mathcal{X}^2$ -distribution with mean $\mu_0=l/N$ and variance $\sigma^2_0=2l/N^2$ for the right key guess, while for the wrong key guess, it follows a $\emph{X}^2$-distribution with mean $\mu_1=l/N+l/2^n$ and standard deviation $\sigma_1=\sqrt{2l}/N+\sqrt{2l}/2^n$. If we denote the probability of false positives and the probability of false
negatives to distinguish between a wrong key and a right key as $\beta_1$ and $\beta_0$,
respectively, and we consider the decision threshold $\tau =\mu_0+\sigma_0z_{1-\beta_0}=\mu_1-\sigma_{1}{z_{1-\beta_1}}$,
then the number of known plaintexts $N$ should be approximately:
$$N=\frac{2^n(z_{1-\beta_0}+z_{1-\beta_1})}{\sqrt{l/2}-z_{1-\beta_1}},\eqno{(1)}$$
where $z_{1-\beta_0}$ and $z_{1-\beta_1}$are the respective quantiles of the standard normal distribution.

Recently, Bogdanov et al. [12] proposed a multidimensional zero-correlation linear distinguisher using $l$ zero-correlation linear approximations to remove the statistical independency assumption, which requires $O(2^n/\sqrt{l})$ known plaintexts, where $n$ is the block size of a cipher.

We treat the zero-correlation linear approximations available as a linear
space spanned by $m$ base zero-correlation linear approximations such that all
$l=2^m -1$ non-zero linear combinations of them have zero correlation. For
each of the $2^m$ data values $z \in F_2^m $, the attacker initializes a counter $V[z]$, $z=0, 1,...,2^m-1$ to value zero.
Then, for each distinct plaintext, the attacker computes the corresponding data value in $F^m_2$ by evaluating the $m$ basis linear
approximations and increments the counter $V[z]$ of this data value by one. Then the attacker computes the statistic $T$:
$$T=\sum_{i=0}^{2^m-1}\frac{(v[z]-N2^{-m})^2}{N2^{-m}(1-2^{-m})}.$$
The statistic $T$  follows a $\mathcal{X}^2$ -distribution with mean $\mu_0=(l-1)\frac{2^n-N}{2^n-1}$ and variance
$\sigma^2_0=2(l-1)\big(\frac{2^n-N}{2^n-1}\big)^2$ for the right key guess, while for the wrong key guess,
it follows a $\emph{X}^2$-distribution with mean $\mu_1=l-1$ and variance $\sigma_1^2=2(l-1)$.

If we denote the probability of false positives and the probability of false negatives to distinguish between a wrong key and a right key as $\beta_0$ and $\beta_1$, respectively, and we consider the decision threshold $\tau =\mu_0+\sigma_0z_{1-\beta_0}=\mu_1-\sigma_{1}{z_{1-\beta_1}}$, then the number of known plaintexts $N$ should be about
$$N=\frac{(2^n-1)(z_{1-\beta_0}+z_{1-\beta_1})}{\sqrt{(l-1)/2}+z_{1-\beta_0}}+1.\eqno{(2)}$$

\section{\large\bf Some zero-correlation linear approximations for 4-round ARIA }
\begin{figure}
 \centering
  \includegraphics[width=9cm]{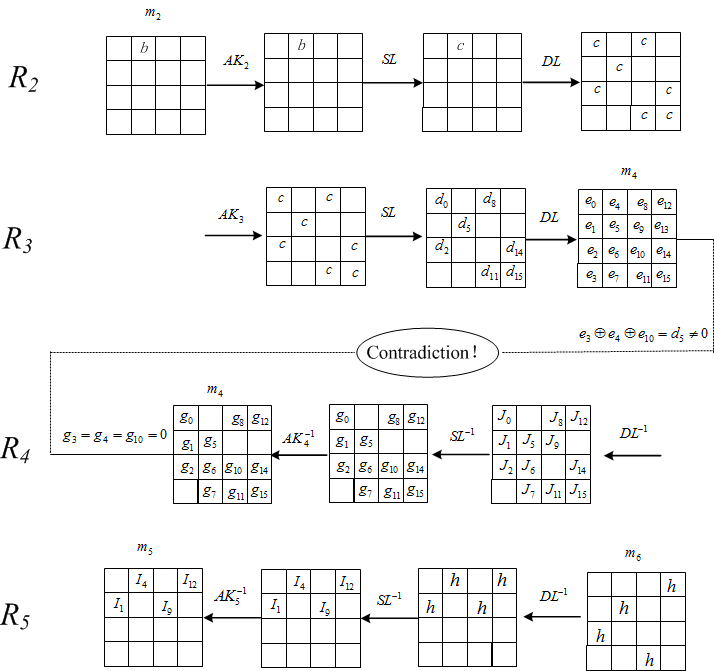}
  \caption{Zero-correlation linear approximations of 4-round ARIA}
\end{figure}
In this section, we show some zero-correlation linear approximations for 4-round ARIA, following the properties on the propagation of linear masks over basic block cipher operations proposed in [10]. We consider the 4-round linear approximations with zero-correlation, which is built in a miss-in-the-middle manner. Some 2-round linear approximations with nonzero bias is concatenated to some 2-round linear approximations with nonzero bias in the inverse direction, where the intermediate masks states contradict with each other.

We assert that the 4-round  linear approximations
$$(0,0,0,0,b,0,0,0; 0,0,0,0,0,0,0,0)\overset{\text{4-Round}}{\longrightarrow}(0,0,h,0,0,h,0,0; 0,0,0,h,h,0,0,0).$$
have zero-correlation, where $b$ and $h$ denote any non-zero value.

Consider that the input masks $(0,0,0,0,b,0,0,0; 0,0,0,0,0,0,0,0)$ will result that the input mask for
 $R_2$ is $(e_0,e_1,...,e_{14},e_{15})$ in the forward direction, where $e_i$, $0 \leq i\leq 15$ denotes a unknown value. The three bytes $e_{3}, e_{4},e_{10}$ satisfy that $e_{3}\oplus e_{4}\oplus e_{10}=d_5 \neq 0$.
In the backward direction, we can get that the input mask of $R_4$ is $(g_0,g_1,...,g_{14},g_{15})$ from the output $(0,0,h,0,0,h,0,0; 0,0,0,h,h,0,0,0)$ where $g_i$, $0 \leq i\leq 15$ also denotes a unknown value. Then, we have $g_{3}=0, g_{4}=0, g_{10}=0$, which leads that $g_{3}\oplus g_{4}\oplus g_{10}=0$ and it contradicts with $d_5 \neq 0$. As a result, the linear hull is  a zero-correlation linear hull, see Figure 2. We also have the following 4-round  linear approximations with zero-correlation,
$$(0,0,0,0,0,0,0,0; 0,0,b,0,0,0,0,0)\overset{\text{4-Round}}{\longrightarrow}(0,0,h,0,0,h,0,0; 0,0,0,h,h,0,0,0);$$
$$(0,0,0,0,0,0,0,0; 0,0,0,0,0,b,0,0)\overset{\text{4-Round}}{\longrightarrow}(0,0,h,0,0,h,0,0; 0,0,0,h,h,0,0,0);$$
$$(0,0,0,0,0,0,0,0; 0,0,0,0,0,0,0,b)\overset{\text{4-Round}}{\longrightarrow}(0,0,h,0,0,h,0,0; 0,0,0,h,h,0,0,0).$$

In addition, the linear map $P$ of diffusion layer is a diagonal matrix. If we treat the input/output masks as the input/output differentials, they are also 4-round impossible differentials.

\section{\large\bf Key-recovery attacks on 6-round ARIA with  Partial-Sum and FFT}

In this section, based on the first 4-round zero-correlation linear approximates, we present some key-recovery attacks on 6-round ARIA-128 with zero-correlation linear cryptanalysis. In the attack process, the Partial-sum and FFT techniques are used to speed up, respectively.

\subsection{\large\bf Key-recovery attacks on 6-round ARIA with Partial-sum technique}

To attack 6-round ARIA, the 4-round linear approximates with zero-correlation start from  round 2 and end
at  round 5. One round is added before and one round is
appended after the linear approximates, refer to Figure 3.
The partial encryption and decryption using the partial sum technique are
proceeded as follows.
\begin{figure}
 \centering
  \includegraphics[width=11cm]{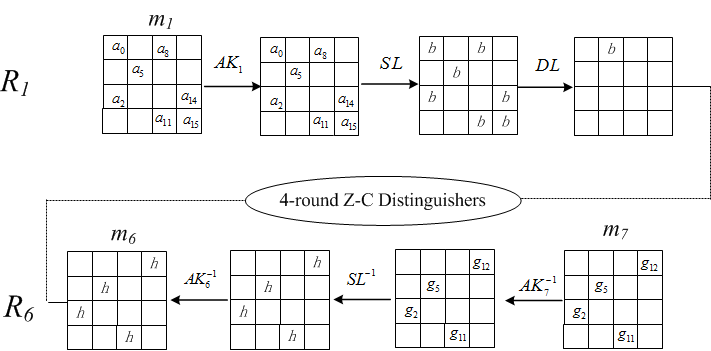}
  \caption{Key-recovery Attack on 6-Round ARIA}
\end{figure}

1.  Allocate 40-bit counters $V_1[x_1]$ for $2^{88}$ possible values of $x_1= m_1[0,2,5,8,11,14,15]|m_7[2,\\ 5,11,12]$ and initialize them to zero. For the corresponding ciphertexts after 6 round encryption, extract the value of $x_1$ and increment the corresponding counter $V_1[x_1]$. The time complexity
of this step is $N$ memory accesses to process the chosen $P C$ pairs. We assume that processing each $P C$ pair is equivalent to one round encryption, then the time complexity of this step is about $N  \times 1/6$ 6-round encryptions.

2. Allocate a counter $V_2[x_2]$ for $2^{80}$ possible values of $x_2= m_1[0,2,5,8,11,14,15]|m_7[5,11,\\12]|I^1_1$ and initialize them to zero.
Guess $k_{7}[2]$ and partially decrypt $x_1$ to get the value of $x_2$, that is, compute $I^1_1= S^{-1}_{6,2}(m_7[2]\oplus k_7[2])$, then update the corresponding counter by
$V_2[x_2] + = V_1[x_1]$. The computation is about $2^{88} \times 2^{8} \times 1/16 \times 1/6 $ 6-round encryptions.

The following steps in the partial encryption and decryption phase are similar
to Step 2, we use Table 2 to show the
details of each partial encryption and decryption step. In Table 2, the second
column stands for the counters should be  allocated in this step. The subkey bytes that have to be guessed in each step are shown in the
third column. the fourth column denotes the time complexity of corresponding step measured in $1/16\times 1/6$
6-round encryption. The intermediate state values are shown in the last column.
\begin{table}[tbp]
\centering
\scriptsize
\begin{tabular}{clllll}
\hline
Step & Counter State & Guess & Complexity & Computed States  \\
\hline
\vspace{-0.1in}3 &$m_1[0,2,5,8,11,14,15]|m_7[11,12]|I^1_2$&$k_7[5]$ &  $2^{88}\times 2^{16} $& $I^1_2=I_1\oplus S^{-1}_{6,5}(m_7[5]\oplus k_7[5])$ \\
\vspace{-0.1in}4 &$m_1[0,2,5,8,11,14,15]|m_7[12]|I^1_3$&$k_7[11]$  &  $2^{80}\times 2^{24} $& $I^1_3=I^1_2\oplus S^{-1}_{6,11}(m_7[11]\oplus k_7[11])$ \\
\vspace{-0.1in}5 &$m_1[0,2,5,8,11,14,15]|I^1_4$ &$k_7[12]$ &  $2^{72}\times 2^{32} $& $I^1_4=I^1_3\oplus S^{-1}_{6,12}(m_7[12]\oplus k_7[12])$ \\
\vspace{-0.1in}6 &$m_1[2,5,8,11,14,15]|I^1_4|I^2_1$&$k_1[0]$ &  $2^{64}\times 2^{40} $& $I^2_1= S_{1,0}(m_1[0]\oplus k_1[0])$ \\
\vspace{-0.1in}7 &$m_1[5,8,11,14,15]|I^1_4|I^2_2$&$k_1[2]$ &  $2^{64}\times 2^{48} $& $I^2_2= I^2_1\oplus S_{1,2}(m_1[2]\oplus k_1[2])$ \\
\vspace{-0.1in}8 &$m_1[8,11,14,15]|I^1_4|I^2_3$&$k_1[5]$ &  $2^{56}\times 2^{56} $& $I^2_3= I^2_2\oplus S_{1,5}(m_1[5]\oplus k_1[5])$ \\
\vspace{-0.1in}9 &$m_1[11,14,15]|I^1_4|I^2_4$&$k_1[8]$ &  $2^{48}\times 2^{64} $& $I^2_4= I^2_3\oplus S_{1,8}(m_1[8]\oplus k_1[8])$ \\
\vspace{-0.1in}10 &$m_1[14,15]|I^1_4|I^2_5$&$k_1[11]$ &  $2^{40}\times 2^{72} $& $I^2_5= I^2_4\oplus S_{1,11}(m_1[11]\oplus k_1[11])$ \\
\vspace{-0.1in}11 &$m_1[15]|I^1_4|I^2_6$&$k_1[14]$ &  $2^{32}\times 2^{80} $& $I^2_6= I^2_5\oplus S_{1,14}(m_1[14]\oplus k_1[14])$ \\
12 &$I^1_4|I^2_7$&$k_1[15]$ &  $2^{24}\times 2^{88} $& $I^2_7= I^2_6\oplus S_{1,15}(m_1[15]\oplus k_1[15])$ \\
\hline
\end{tabular}

\caption{Partial Encryption and Decryption of the Attack on 6-Round ARIA-128}
\end{table}

13. Allocate a counter vector $V[z]$ of size $2^{16}$ where each
element is 120-bit length for 16-bit z (z is the concatenation of evaluations of 16 basis zero-correlation masks). For $2^{16}$ values of $x_{12}$, evaluate all basis zero-correlation masks on $V_{12}$ and put the evaluations to the vector $z$, then add the corresponding $V [z]: V [z]+ = V_{12}[x_{12}]$. Compute $T = N 2^{16}\sum_{z=0}^{2^{16}-1}(\frac{v[z]}{N}-\frac{1}{2^{16}})$, if $T < \tau$ , then the guessed key is a possible key candidate.

In this attack, we set the type-I error probability $\beta_0 =2^{-2.7}$ and the type-II error probability $\beta_1 =2^{-90}$. We have
$z_{1-\beta_0}\approx 1$, $z_{1-\beta_1}\approx 11$, $n = 128$, $ l=2^{16}$. According to Equation (2) The date complex $N$ is about $2^{123.6}$ and the decision threshold $\tau \approx 2^{15.9}$.

There are 88-bit key values guessed during the encryption phase, and only the right key candidates survive in the wrong key filtration. The complexity of Step 3 to Step 12 is no more than $2^{108.6}$ 6-round ARIA encryptions and  the complexity of Step 1 is about $2^{121}$ 6-round ARIA encryptions which is also the dominant part of our attack. In total, the data complexity is about $2^{123.6}$ known plaintexts, the time complexity is about $2^{121}$ 6-round ARIA encryptions and the memory requirement are about $2^{90.3}$ byte for counters.

\subsection{\bf Key-recovery attack on 6-round ARIA  with FFT technique}
Using the FFT technique, we can attack 6-round AIRA-128 starting from the first round by placing the 4-round zero-correlation
linear approximations in rounds 2 to 5. One round is added before and one round is
appended after the linear approximates, also see  Figure 3.

In our attack, we guess the subkey and evaluate the linear approximation
$(0,0,0,0,b,0,0,0; \\0,0,0,0,0,0,0,0) \cdot m_2
\oplus (0,0,h,0,0,h,0,0; 0,0,0,h,h,0,0,0) \cdot m_6 = 0 $, that is ,
$$
u = b\cdot \big(\oplus_{i=0,2,5,8,11,14,15}S_{1,i}(m_1[i]\oplus k_1[i])\big)
\oplus h \cdot \big(\oplus_{i=2,5,11,12}S^{-1}_{6,i}(m_7[i]\oplus k_7[i])\oplus k_6[i]\big)=0.
$$

Let $k_{6}=k_6[2]\oplus k_6[5]\oplus k_6[11]\oplus k_6[13]$ and $v=u\oplus b\cdot k_{6}$, then we have
$$v = b\cdot \big(\oplus_{i=0,2,5,8,11,14,15}S_{1,i}(m_1[i]\oplus k_1[i])\big)
\oplus h \cdot \big(\oplus_{i=2,5,11,12}S^{-1}_{6,i}(m_7[i]\oplus k_7[i])\big)=0.
\eqno{(3)}$$

Our attack is equivalent to evaluating the
correlation of the linear approximation $v = 0$. The correlation
of the linear approximation $v = 0$ can be evaluated as the matrix vector product where
the matrix is:
$$ M(m_1[0,2,5,8,11,14,15]|m_7[2,5,11,12]|k_1[0,2,5,8,11,14,15]|k_7[2,5,11,12])={(-1)}^{v},
\eqno{(4)}$$
see [13] and [18] for detail.  Then the attack is performed as follows:

1. Allocate the vector of counters $V_K$ of the experimental correlation for every
subkey candidate $K = k_1[0,2,5,8,11,14,15]|k_7[2,5,11,12]$.

2. For each of $N$ plaintext-ciphertext pairs, extract the 88-bit value
$i= m_1[0,2,5,8,11,14,\\15] |m_7[2,5,11,12]$ and increment the counters $x_i$ according to the value of $i$.

3. For each of the $2^{16}$ linear approximations,

  (i). Perform the key counting phase and compute the first column of $M$ using (3) and (4). As $M$ is a 88-level circulant matrix, this information is sufficient to define $M$ completely , which requires $2^{88}$ operations.

 (ii). Evaluate the vector $\epsilon = M \cdot x $, which requires about $3 \times 88 \times 2^{88} $ operations.

(iii). Let $W=W+(\epsilon/N )^2$, If $W < \tau$, then the corresponding $K$ is a possible subkey candidate and all
master keys are tested exhaustively.

After Step 3, we obtain $2^{88}$ counters $V_K$ which are the sum of squares of
correlations for $2^{16}$ linear approximations under each $k$. The correct subkey is
then selected from the candidates with $V_K$ less than the threshold  $\tau$
If we set $\beta_0 = 2^{-2.7}$ and $\beta_1 = 2^{-90}$, we get $z_{1-\beta_0}\approx 1$ and $z_{1-\beta_1}\approx 11$. Since
the block size $n = 128$ and we have $l = 2^{16}$ linear approximations, according to
Equation (1), the number of known plaintext-ciphertext pairs $N$ should be about
$2^{124.1}$ and the threshold  $\tau\approx 2^{-108.4}$.
In Step 3, only the right guess is expected to survive for the 88-bit target subkey. The complexities for Step 2, Step 3, are $2^{121.5}$ memory accesses,  $2^{16}\times 4 \times 88\times 2^{88} = 2^{112.5}$ operators, respectively. If we assume that one time of memory access, one time of operators, one 6-round Camellia encryption  are equivalent, then the total time
complexity is about $2^{121.5}$ encryptions. The memory requirements are about $2^{90.3}$
bytes.


\section{\large\bf Key-recovery attacks on 7-round ARIA with  Partial-Sum and FFT}

In this section, we describe some zero-correlation linear cryptanalysis of 7-round ARIA. The attack is based on the first 4-round zero-correlation linear approximates with additional one round in the begin and two rounds at the end, see Fig.4. Partial-Sum and FFT are also used in the attack process, respectively.
\subsection{\large\bf Key-recovery attacks on 7-round ARIA with Partial-sum technique}

Similarly to the attacks to 6-round ARIA, the partial encryption and decryption using the partial sum technique are
proceeded as follows.
\begin{figure}
 \centering
  \includegraphics[width=10cm]{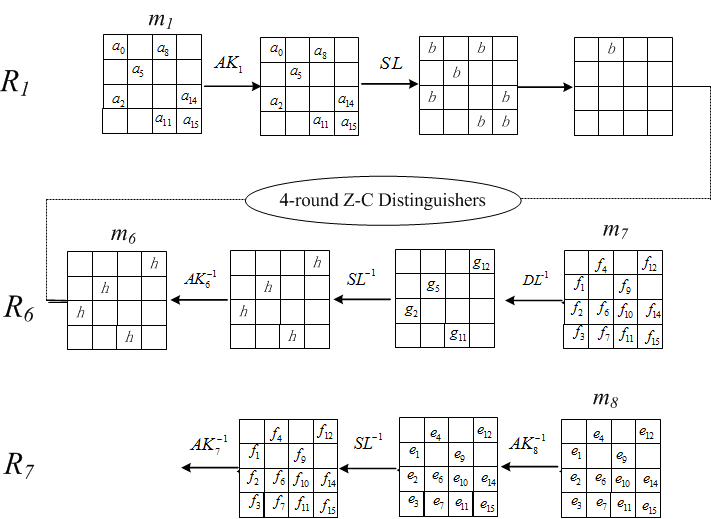}
  \caption{Key-recovery Attack on 7-Round ARIA}
\end{figure}

1.  Allocate 8-bit counters $V_1[x_1]$ for $2^{152}$ possible values of $x_1= m_1[0,2,5,8,11,14,15]|m_8[1,\\ 2,3,4,6,7,9,10,11,12,14,15]$ and initialize them to zero. For the corresponding ciphertexts after 7 round encryption, extract the value of $x_1$ and increment the corresponding counter $V_1[x_1]$. The time complexity
of this step is $N$ memory accesses to process the chosen $P C$ pairs. We assume that processing each $P C$ pair is equivalent to one round encryption, then the time complexity of this step is about $N  \times 1/7$ 7-round encryptions.

2. Allocate 8-bit counters $V_2[x_2]$ for $2^{120}$ possible values of $x_2= m_1[2,5,8,11,14,15]|m_8[11,\\ 12,14,15]|I^1_1|I^2_1|I^5_1|I^{11}_1|I^{12}_1$ and initialize them to zero.
Guess $k_{1}[0]$,$k_{8}[1,2,3,4,6,7,9,10]$ and partially decrypt $x_1$ to get the value of $x_2$, that is, compute
$I^1_1= S_{1,0}(m_1[0]\oplus k_1[0])$,
 $I^2_1= \oplus_{i=1,4,6,10} S^{-1}_{7,i}(m_8[i]\oplus k_8[i])$,
$I^5_1= \oplus_{i=1,3,4,9,10} S^{-1}_{7,i}(m_8[i]\oplus k_8[i])$,
$I^{11}_1= \oplus_{i=2,3,4,7,9} S^{-1}_{7,i}(m_8[i]\oplus k_8[i])$,
$I^{12}_1= \oplus_{i=1,2,6,7,9} S^{-1}_{7,i}(m_8[i]\oplus k_8[i])$, then update the corresponding counter by
$V_2[x_2] + = V_1[x_1]$. The computation in this step  is no more than $N \times 2^{72} \times 1/16 \times 1/7 $ 7-round encryptions.

\begin{table}[tbp]
\centering
\scriptsize
\begin{tabular}{clllll}
\hline
Step & Counter State & Guess & Complexity & Computed States  \\
\hline
\vspace{-0.1in}7 &$m_1[5,8,11,14,15]|I^1_2|I^2_4|I^5_3|I^{11}_3|I^{12}_3$&$k_1[2]$ &  $2^{88}\times 2^{112} $& $I^1_2=I^1_1\oplus S_{1,2}(m_1[2]\oplus k_1[2])$ \\
\vspace{-0.1in}8 &$m_1[8,11,14,15]|I^1_3|I^2_4|I^5_3|I^{11}_3|I^{12}_3$&$k_1[5]$ &  $2^{80}\times 2^{120} $& $I^1_3=I^1_2\oplus S_{1,5}(m_1[5]\oplus k_1[5])$ \\
\vspace{-0.1in}9 &$m_1[11,14,15]|I^1_4|I^2_4|I^5_3|I^{11}_3|I^{12}_3$&$k_1[8]$ &  $2^{72}\times 2^{128} $& $I^1_4=I^1_3\oplus S_{1,8}(m_1[8]\oplus k_1[8])$ \\
\vspace{-0.1in}10 &$m_1[14,15]|I^1_5|I^2_4|I^5_3|I^{11}_3|I^{12}_3$&$k_1[11]$ &  $2^{64}\times 2^{136} $& $I^1_5=I^1_4\oplus S_{1,11}(m_1[11]\oplus k_1[11])$ \\
\vspace{-0.1in}11 &$m_1[15]|I^1_6|I^2_4|I^5_3|I^{11}_3|I^{12}_3$&$k_1[14]$ &  $2^{56}\times 2^{144} $& $I^1_6=I^1_5\oplus S_{1,14}(m_1[14]\oplus k_1[14])$ \\
\vspace{-0.1in}12 &$I^1|I^2_4|I^5_3|I^{11}_3|I^{12}_3$&$k_1[15]$ &  $2^{48}\times 2^{152} $& $I^1=I^1_6\oplus S_{1,15}(m_1[15]\oplus k_1[15])$ \\
\vspace{-0.1in}13 &$I^1|I^2|I^5_3|I^{11}_3|I^{12}_3$&$k_{7,2}$ &  $2^{48}\times 2^{160} $& $I^2=S^{-1}_{6,2}(I^2_4\oplus k_{7,2})$ \\

\vspace{-0.1in}14 &$I^1|I^5|I^{11}_3|I^{12}_3$&$k_{7,5}$ &  $2^{40}\times 2^{168} $& $I^5=I^2\oplus S^{-1}_{6,5}(I^5_3\oplus k_{7,5})$ \\

\vspace{-0.1in}15 &$I^1|I^{11}|I^{12}_3$&$k_{7,11}$ &  $2^{32}\times 2^{176} $& $I^{11}=I^5\oplus S^{-1}_{6,11}(I^{11}_3\oplus k_{7,11})$ \\

16 &$I^1|I^{12}$&$k_{7,12}$ &  $2^{24}\times 2^{184} $& $I^{12}=I^{11}\oplus S^{-1}_{6,12}(I^{12}_3\oplus k_{7,12})$ \\

\hline
\end{tabular}

\caption{Partial Encryption and Decryption of the Attack on 7-Round ARIA}
\end{table}

3. Allocate 8-bit counters $V_3[x_3]$ for $2^{112}$ possible values of $x_3= m_1[2,5,8,11,14,15]|m_8[12,\\ 14,15]|I^1_1|I^2_2|I^5_1|I^{11}_1|I^{12}_2$ and initialize them to zero.
Guess $k_{8}[11]$ and partially decrypt $x_2$ to get the value of $x_3$, that is, compute
 $I^2_2= I^2_1\oplus S^{-1}_{7,11}(m_8[11]\oplus k_8[11])$,
$I^{12}_2= I^{12}_1\oplus S^{-1}_{7,11}(m_8[11]\oplus k_8[11])$, then update the corresponding counter by
$V_3[x_3] + = V_2[x_2]$. The computation in this step  is no more than $2^{120} \times 2^{80} \times 1/16 \times 1/7 $ 7-round encryptions.

4. Allocate 16-bit counters $V_4[x_4]$ for $2^{104}$ possible values of $x_4= m_1[2,5,8,11,14,15]|m_8[14,\\ 15]|I^1_1|I^2_3|I^5_1|I^{11}_2|I^{12}_3$ and initialize them to zero.
Guess $k_{8}[12]$ and partially decrypt $x_3$ to get the value of $x_4$, that is, compute
 $I^2_3= I^2_2\oplus S^{-1}_{7,12}(m_8[12]\oplus k_8[12])$,
 $I^{11}_2= I^{11}_1\oplus S^{-1}_{7,12}(m_8[12]\oplus k_8[12])$,
$I^{12}_3= I^{12}_2\oplus S^{-1}_{7,12}(m_8[12]\oplus k_8[12])$, then update the corresponding counter by
$V_4[x_4] + = V_3[x_3]$. The computation in this step  is no more than $2^{112} \times 2^{88} \times 1/16 \times 1/7 $ 7-round encryptions.

5. Allocate 24-bit counters $V_5[x_5]$ for $2^{96}$ possible values of $x_5= m_1[2,5,8,11,14,15]|m_8[15]\\|I^1_1|I^2_3|I^5_1|I^{11}_2|I^{12}_3$ and initialize them to zero.
Guess $k_{8}[14]$ and partially decrypt $x_4$ to get the value of $x_5$, that is, compute
 $I^5_2= I^5_1\oplus S^{-1}_{7,14}(m_8[14]\oplus k_8[14])$,
 $I^{11}_3= I^{11}_2\oplus S^{-1}_{7,14}(m_8[14]\oplus k_8[14])$,
 then update the corresponding counter by
$V_5[x_5] + = V_4[x_4]$. The computation in this step  is no more than $2^{104} \times 2^{96} \times 1/16 \times 1/7 $ 7-round encryptions.

6. Allocate 32-bit counters $V_6[x_6]$ for $2^{88}$ possible values of $x_6= m_1[2,5,8,11,14,15]|I^1_1\\|I^2_3|I^5_2|I^{11}_3|I^{12}_3$ and initialize them to zero.
Guess $k_{8}[15]$ and partially decrypt $x_5$ to get the value of $x_6$, that is, compute
 $I^2_4= I^2_3\oplus S^{-1}_{7,15}(m_8[15]\oplus k_8[15])$,
 $I^{5}_3= I^{5}_2\oplus S^{-1}_{7,15}(m_8[15]\oplus k_8[15])$,
 then update the corresponding counter by
$V_6[x_6] + = V_5[x_5]$. The computation in this step  is no more than $2^{96} \times 2^{104} \times 1/16 \times 1/7 $ 7-round encryptions.

 Similarly, we use Table 3 to show the
details of each partial encryption and decryption step, where we let $k_{7,2}=\oplus_{i=1,4,6,10,11,12,15}k_{7}[i]$,
$k_{7,5}=\oplus_{i=1,3,4,9,10,14,15}k_{7}[i]$, $k_{7,11}=\oplus_{i=2,3,4,7,9,12,14}k_{7}[i]$ and $k_{7,12}=\oplus_{i=1,2,6,7,9,11,12}k_{7}[i]$.
After Step 13, we have reached the boundaries of the zero-correlation linear
approximations over 7-round ARIA. We then proceed the following steps to recover
the right key.

%
%
%

17. Allocate a counter vector $V[z]$ of size $2^{16}$ where each
element is 120-bit length for 16-bit z (z is the concatenation of evaluations of 16 basis zero-correlation masks). For $2^{16}$ values of $x_{16}$, evaluate all basis zero-correlation masks on $V_{16}$ and put the evaluations to the vector $z$, then add the corresponding $V [z]: V [z]+ = V_{16}[x_{16}]$. Compute $T = N 2^{16}\sum_{z=0}^{2^{16}-1}(\frac{v[z]}{N}-\frac{1}{2^{16}})$, if $T <\tau$ ,then the guessed key is a possible key candidate. We do exhaustive search for all keys conforming to this possible key candidate.

In this attack, we set the type-I error probability $\beta_0 =2^{-2.7}$ and the type-II error probability $\beta_1 =2^{-186}$. We have
$z_{1-\beta_0}\approx 1$, $z_{1-\beta_1}\approx 15.7$, $n = 128$, $ l=2^{16}$. According to Equation (2), the date complex $N$ is about $2^{124.6}$ and the decision threshold $\tau \approx 2^{15.9}$.

There are 184-bit key values guessed during the encryption phase, and only the right key candidates survive in the wrong key filtration. The complexity of Step 3 to Step 16 is no more than $2^{203.5}$ 7-round ARIA encryptions and   In total, the data complexity is about $2^{124.6}$ known plaintexts, the time complexity is about $2^{203.5}$ 7-round ARIA encryptions and the memory requirement are about $2^{152}$ byte for counters.

\subsection{\bf Key-recovery attack on 7-round ARIA  with FFT technique}

In our attack, one round is added before and two rounds are
appended after the linear approximates with zero-correlation from rounds 2 to 5, see  Figure 4. We evaluate the linear approximations
$(0,0,0,0,b,0,0,0; 0,0,0,0,0,0,0,0) \cdot m_2
\oplus (0,0,h,0,0,h,0,0; 0,0,0,h,h,0,0,0) \cdot m_6 = 0 $, that is ,
\begin{equation*}
\begin{aligned}
u = b\cdot \big(&\oplus_{i=0,2,5,8,11,14,15}S_{1,i}(m_1[i]\oplus K_1[i])\big)
\oplus h \cdot \big(S^{-1}_{6,2}\big(\oplus_{i=1,4,6,10,11,12,15}\big(S^{-1}_{7,i}(m_8[i]\oplus k_8[i])\\ \oplus& k_7[i]\big)\big)
\oplus S^{-1}_{6,5}\big(\oplus_{i=1,3,4,9,10,14,15}(S^{-1}_{7,i}(m_8[i]\oplus k_8[i])\oplus k_7[i])\big)
\\&\oplus S^{-1}_{6,11}\big(\oplus_{i=2,3,4,7,9,12,14}(S^{-1}_{7,i}(
m_8[i]\oplus k_8[i])\oplus k_7[i])\big)\\&
\oplus S^{-1}_{6,12}\big(\oplus_{i=1,2,6,7,9,11,12}(S^{-1}_{7,i}(m_8[i]\oplus k_8[i])\oplus k_7[i])\big)\\
&\oplus (k_6[2]\oplus k_6[5]\oplus k_6[11]\oplus k_6[12])\big)=0.
\end{aligned}
\end{equation*}

Let  $K_{7,2}=\oplus_{i=0,2,5,8,11,14,15}K_7[i]$,
$K_{7,5}=\oplus_{i=1,4,6,10,11,12,15}K_7[i]$,
$K_{7,11}=\oplus_{i=1,3,4,9,10,14,15}\\ K_7[i]$,
$K_{7,12}=\oplus_{i=2,3,4,7,9,12,14}K_7[i]$,
 $K_{6}=k_6[2]\oplus k_6[5]\oplus k_6[11]\oplus k_6[12]$, and $v=u\oplus b\cdot K_{6}$, then we have
\begin{equation*}
\begin{aligned}
v = b\cdot \big(&\oplus_{i=1,6,8,10,13,15}S_{1,i}(m_1[i]\oplus K_1[i])\big)
\oplus h \cdot \big(S^{-1}_{6,2}\big(\oplus_{i=0,2,5,8,11,14,15}S_{8,i}^{-1}(m_7[i]\\ & \oplus K_8[i]) \oplus K_{7,3}\big)
\oplus S^{-1}_{6,5}(\oplus_{i=1,3,4,9,10,14,15}S_{7,i}^{-1}(m_8[i]\oplus k_8[i])\oplus k_{7,5})\\
&\oplus S^{-1}_{6,11}(\oplus_{i=2,3,4,7,9,12,14}S_{7,i}^{-1}(m_8[i]\oplus k_8[i])\oplus k_{7,11})\\
&\oplus S^{-1}_{6,12}(\oplus_{i=1,2,6,7,9,11,12}S_{7,i}^{-1}(m_8[i]\oplus k_8[i])\oplus k_{7,12})\big)=0.
\end{aligned}\eqno{(5)}
\end{equation*}

Our attack is equivalent to evaluating the
correlation of the linear approximation $v = 0$, which can be evaluated as the matrix vector product where the matrix is:

\begin{equation*}
\begin{aligned}
M\big(m_1&[0,2,5,8,11,14,15]|m_8[1,2,3,4,6,7,9,10,11,12,14,15]|k_1[0,2,5,8,11,14,15]\\&|k_8[1,2,3,4,6,7,9,10,11,12,14,15]| k_{7,2}|k_{7,5}| k_{7,11}| k_{7,12}\big) = {(-1)}^{v}.
\end{aligned}\eqno{(6)}
\end{equation*}
 Then the attack is performed as follows:

1. Allocate the vector of counters $V_K$ of the experimental correlation for every
subkey candidate $K = k_1[0,2,5,8,11,14,15]|k_8[1,2,3,4,6,7,9,10,11,12,14,15]| k_{7,2}|k_{7,5}| k_{7,11}| k_{7,12}$.

2. For each of $N$ plaintext-ciphertext pairs, extract the 8-bit value
$i= m_1[0,2,5,8,11,14,15]\\ |m_8[1,2,3,4,6,7,9,10,11,12,14,15]$, increment the counters $x_i$ according to the value of $i$.

3. For each of the $2^{16}$ linear approximations,

  (i). Perform the key counting phase and compute the first column of $M$ using (5) and (6). As $M$ is a 184-level circulant matrix, this information is sufficient to define $M$ completely , which requires $2^{184}$ operations.

 (ii). Evaluate the vector $\epsilon = M \cdot x $, which requires about $3 \times 184 \times 2^{184} $ operations.

(iii). Let $W=W+(\epsilon/N )^2$, If $W < \tau$, then the corresponding $K$ is a possible subkey candidate and all
master keys are tested exhaustively.

After Step 3, we obtain $2^{184}$ counters $V_K$ which are the sum of squares of
correlations for $2^{16}$ linear approximations under each $k$. The correct subkey is
then selected from the candidates with $V_K$ less than the threshold  $\tau$
If we set $\beta_0 = 2^{-2.7}$ and $\beta_1 = 2^{-186}$, we get $z_{1-\beta_0}\approx 1$ and $z_{1-\beta_1}\approx 15.7$. Since
the block size $n = 128$ and we have $l = 2^{16}$ linear approximations, according to
Equation (1), the number of known plaintext-ciphertext pairs $N$ should be about
$2^{124.7}$ and the threshold  $\tau\approx 2^{-108.4}$.
In Step 3, only the right guess is expected to survive for the 184-bit target subkey. The complexities for Step 2, Step 3, are $2^{121.9}$ memory accesses,  $2^{16}\times 4 \times 184 \times 2^{184} = 2^{209.5}$ operators, respectively. If we assume that one time of memory access, one time of operators, one 7-round Camellia encryption  are equivalent, then the total time
complexity is about $2^{209.5}$ encryptions. The memory requirements are about $2^{152}$ bytes.


\section{\large\bf Conclusion }
In this paper, we evaluate the security of ARIA block cipher with respect to the technique of zero-correlation linear  cryptanalysis.
 We deduce some 4-round  zero-correlation linear approximations of ARIA, and based on those linear approximations, we give some key-recovery
 attacks on  6/7 round  ARIA-128/256  with the Partial-sum technique and FFT technique taken into consideration. For the first time, we consider the security of ARIA against zero-correlation linear cryptanalysis. While two techniques are used in the attack, it also gives us a chance to compare the partial-sum technique and the FFT technique.

\vspace{0.5in}

\leftline{\bf References} \bigbreak
\def\REF#1{\par\hangindent\parindent\indent\llap{#1\enspace}\ignorespaces}
\footnotesize
\small
\REF{[1]} Daesung, K., Jaesung, K., Sangwoo, P., et al. New Block Cipher: ARIA. Information Security and Cryptology-ICISC'03, LNCS, Vol.2971, 2003, pp.432-445.
\REF{[2]} Biryukov, A., Canniere, C.,  et al. Security and Performance Analysis of ARIA. Version 1.2. Jan 7, 2004.
\REF{[3]} Wu, W., Zhang, W., Feng, D. Impossible Differential Cryptanalysis of Reduced-Round ARIA and Camellia. Journal of Computer Science and Technology, Vol.22, 2007, pp. 449-456.
\REF{[4]}Li, S., Song, C. Improved Impossible Differential Cryptanalysis of ARIA. IEEE Computer Society, ISA 2008, pp.129-132.
\REF{[5]}Fleischmann, E., Gorski, M.,  Lucks, S. Attacking Reduced Rounds of the ARIA Block Cipher. http://eprint.iacr.org/2009/334.pdf.2009.07.
\REF{[6]}Li, Y., Wu, W. Zhang, L. Integral Attacks on Reduced-round ARIA Block Cipher. ISPEC,  LNCS, Vol.6047, 2010, pp.19-29.
\REF{[7]}Tang, X., Sun, B.,  Li, R. A Meet-in-the-middle Attack on Reduced-Round ARIA. Journal of Systems and Software, Vol.84, 2011, pp.1685-1692.
\REF{[8]}Du, C., Chen, J. Impossible Differential Cryptanalysis of ARIA Reduced to 7 Rounds. CANS, LNCS, 2010, Vol.6467, pp.20-30.
\REF{[9]} Xie, Z., Chen, S. Impossible Differential Cryptanalysis of 7-Round ARIA-192, Journal of Electronics Information Technology,  Vol.35, 2013, pp.2301-2306.
\REF{[10]}Bogdanov, A., Rijmen, V., Linear Hulls with Correlation Zero and Linear Cryptanalysis of Block Ciphers. Designs, Codes and Cryptography March, Vol.70, 2014, pp.369-383.
\REF{[11]} Bogdanov, A., Wang, M.: Zero Correlation Linear Cryptanalysis with Reduced Data Complexity, FSE 2012, LNCS, vol. 7549, 2012, pp.29-48.
\REF{[12]} Bogdanov, A.,  Leander, G.,   Nyberg, K.,  Wang, M. Integral and Multidimensional Linear Distinguishers with Correlation Zero.  ASIACRYPT 2012, LNCS,  Vol. 7658,  2012, pp.244-261.
\REF{[13]} Bogdanov, A., Geng, H., Wang, M.,  Wen, L., Collard, B. Zero-correlation Linear Cryptanalysis with FFT and Improved Attacks on ISO Standards Camellia and CLEFIA,  SAC¡¯13, LNCS, 2014, pp. 306-323.
\REF{[14]}Wen, L., Wang, M., Bogdanov, A., Chen, H. Multidimensional Zero-Correlation Attacks on Lightweight Block Cipher HIGHT: Improved Cryptanalysis of an ISO Standard. Information Processing Letters, Vol.114, 2014, pp.322-330.
\REF{[15]} Soleimany, H., Nyberg, K. Zero-correlation Linear Cryptanalysis of Reduced-round LBlock. IACR Cryptology ePrint Archive,
http://eprint.iacr.org/2012/570.pdf, Nov. 2013.
\REF{[16]}Wen, L.,Wang, M., Bogdanov, A. Multidimensional Zero-Correlation Linear Cryptanalysis of E2. Progress in Cryptology - AFRICACRYPT 2014, LNCS, Vol. 8469, 2014, pp.147-164.
\REF{[17]} Ferguson, N., Kelsey, J., Lucks, S., Schneier, B., Stay, M., Wagner, D., Whiting, D.: Improved Cryptanalysis of Rijndael. FSE. LNCS, vol.1978, 2000, pp.213-230
\REF{[18]}   Collard, B.,  Standaert, F.,  Quisquater, J. Improving the Time Complexity of Matsui's Linear Cryptanalysis. ICISC 2007, LNCS, 2007, vol. 4817, pp.77-88.
\end{document}